\documentclass[sigconf]{acmart}

\usepackage{subfiles}
\usepackage{makecell}
\usepackage{multirow}
\usepackage{longtable}
\usepackage{graphicx}
\usepackage{caption}
\usepackage{subcaption}
\usepackage[dvipsnames,table]{xcolor}
\usepackage{tabularx}
\usepackage{soul}
\usepackage{float}
\usepackage{url}
\usepackage{booktabs}
\usepackage{colortbl}

\usepackage{enumitem}

\usepackage{tcolorbox}
\tcbuselibrary{skins,breakable}

\newif\ifredact
\redactfalse  

\newcommand{\sk}{\textsc{SketchLLM}}
\newcommand{\dquote}[1]{\textit{``#1''}}

\newif\ifcomment
\commenttrue   
\ifcomment
  \newcommand{\missing}[1]{\textcolor{red}{~#1}}
  \newcommand{\wei}[1]{~\sethlcolor{cyan!40}\hl{[Weiyan: #1]}}
  \newcommand{\ken}[1]{~\sethlcolor{yellow!40}\hl{[Kenny: #1]}}
  \newcommand{\rrev}[3]{\textcolor{blue}{[RevID #1] #2: #3}} 
\else
  \newcommand{\missing}[1]{}
  \newcommand{\wei}[1]{}
  \newcommand{\ken}[1]{}
  \newcommand{\rrev}[3]{}
\fi

\AtBeginDocument{%
  }

\copyrightyear{2026}
\acmYear{2026}
\setcopyright{cc}
\setcctype{by}
\acmConference[CHI EA '26]{Extended Abstracts of the 2026 CHI Conference on Human Factors in Computing Systems}{April 13--17, 2026}{Barcelona, Spain}
\acmBooktitle{Extended Abstracts of the 2026 CHI Conference on Human Factors in Computing Systems (CHI EA '26), April 13--17, 2026, Barcelona, Spain}
\acmDOI{10.1145/3772363.3798524}
\acmISBN{979-8-4007-2281-3/2026/04}

\begin{document}

\title[A Taxonomy of Human--MLLM Interaction]{
A Taxonomy of Human--MLLM Interaction in Early-Stage Sketch-Based Design Ideation}
\author{Weiyan Shi}
\email{weiyanshi6@gmail.com}
\orcid{0009-0001-6035-9678}
\affiliation{
  \institution{Singapore University of Technology and Design}
  \country{Singapore}
  \country{Singapore}
}

\author{Kenny Tsu Wei Choo}
\email{kennytwchoo@gmail.com}
\orcid{0000-0003-3845-9143}
\affiliation{
  \institution{Singapore University of Technology and Design}
  \city{Singapore}
  \country{Singapore}
}

\ccsdesc[500]{Human-centered computing~Empirical studies in HCI}
\ccsdesc[500]{Human-centered computing~Human computer interaction (HCI)}

\keywords{Multimodal large language model, Sketching, Human-AI interaction, Design ideation, Generative AI}


\begin{abstract}
As multimodal large language models (MLLMs) are increasingly integrated into early-stage design tools, it is important to understand how designers collaborate with AI during ideation.
In a user study with 12 participants, we analysed sketch-based design interactions with an MLLM-powered system using automatically recorded interaction logs and post-task interviews.
Based on how creative responsibility was allocated between humans and the AI, we predefined four interaction modes—Human-Only, Human-Lead, AI-Lead, and Co-Evolution—and analysed how these modes manifested during sketch-based design ideation.
Our results show that designers rarely rely on a single mode; instead, human- and AI-led roles are frequently interwoven and shift across ideation instances.
These findings provide an empirical basis for future work to investigate why designers shift roles with AI and how interactive systems can better support such dynamic collaboration.
\end{abstract}

\maketitle

\section{Introduction}
Sketching is a foundational activity in early-stage design ideation, supporting exploration, reflection, and the iterative development of ideas~\cite{landay1995interactive, buxton2010sketching, suwa2022roles}.
Through sketching, designers externalise tentative concepts, reinterpret visual cues, and progressively refine design directions during open-ended creative work.

Recent advances in multimodal large language models (MLLMs) have led to their increasing integration into sketch-based design tools~\cite{fui2023generative, zhang2023generative, davis2025sketchai, lin2025inkspire}.
By enabling models to interpret sketches, generate images, and respond to textual input, these systems position AI as an active participant in early-stage ideation~\cite{vinker2025sketchagent, xu2024llm}.

As multimodal large language models (MLLMs) become embedded in early-stage design workflows, an important open question is how designers actually interact with AI during sketch-based ideation.
While prior work has demonstrated the potential of AI to support creative exploration and ideation~\cite{wang2025aideation, lin2025inkspire, davis2025sketchai, shi2025talksketch}, there remains limited empirical understanding of the interactional structure of human–MLLM collaboration in practice.
In particular, it is unclear how creative responsibility is distributed between humans and AI during ideation, and how this distribution evolves as ideas develop over time.

During sketch-based design ideation, designers may work independently, guide AI behaviour, respond to AI-generated outputs, or collaboratively develop ideas with AI across successive interaction turns~\cite{lawton2023drawing, tholander2023design, shaer2024ai}.
Rather than following a single, fixed interaction pattern, human–AI collaboration is likely to involve multiple modes that emerge, coexist, and shift over time.
However, these modes have rarely been systematically characterised in sketch-based design contexts, particularly with respect to how creative roles are allocated and transition during ongoing ideation.

To address this gap, we investigate human–MLLM interaction in early-stage sketch-based design ideation through a taxonomy-oriented analysis.
Based on a user study with 12 participants, we analyse sketch-based interactions with an MLLM-powered system using automatically recorded interaction logs and post-task interviews~\cite{wright1991use}.
By examining how creative responsibility is allocated between humans and AI, we identify distinct interaction modes and analyse broader trends in how designers transition between them during ideation.
Our goal is to provide empirical insights that can inform the design of more human-centred AI systems that better support dynamic and evolving forms of creative collaboration~\cite{shneiderman2022human}.

We aim to explore the following research questions:

\textit{\textbf{RQ1.}}
How do participants interact with MLLMs during early-stage sketch-based design ideation?

\textit{\textbf{RQ2.}}
When these interactions are categorised based on the allocation of creative responsibility between humans and AI, what patterns and outcomes emerge across the ideation process?
\section{User Study}
\label{sec:userstudy}

We conducted a user study ($N = 12$) to examine how designers complete an early-stage design ideation task with the assistance of a MLLM.
Participants were asked to develop a design concept through sketching while interacting with the AI system, which provided generative support during the task.
The study focused on capturing participants’ interaction behaviours and use patterns when working with an AI-assisted design environment.

This study was reviewed and approved by the Institutional Review Board of Singapore University of Technology and Design (IRB approval no. IRB-25-00717), and all participants provided informed consent. Audio and screen recordings were transcribed and manually verified, with all personally identifiable information removed during processing. Data were stored in encrypted, password-protected institutional systems accessible only to the research team. Recordings and transcripts were not shared with participants. Identifiable data will be securely deleted in accordance with institutional policy.

\subsection{Study Prototype: \sk{}}

\begin{figure*}[t]
\centering
\includegraphics[width=0.8\textwidth]{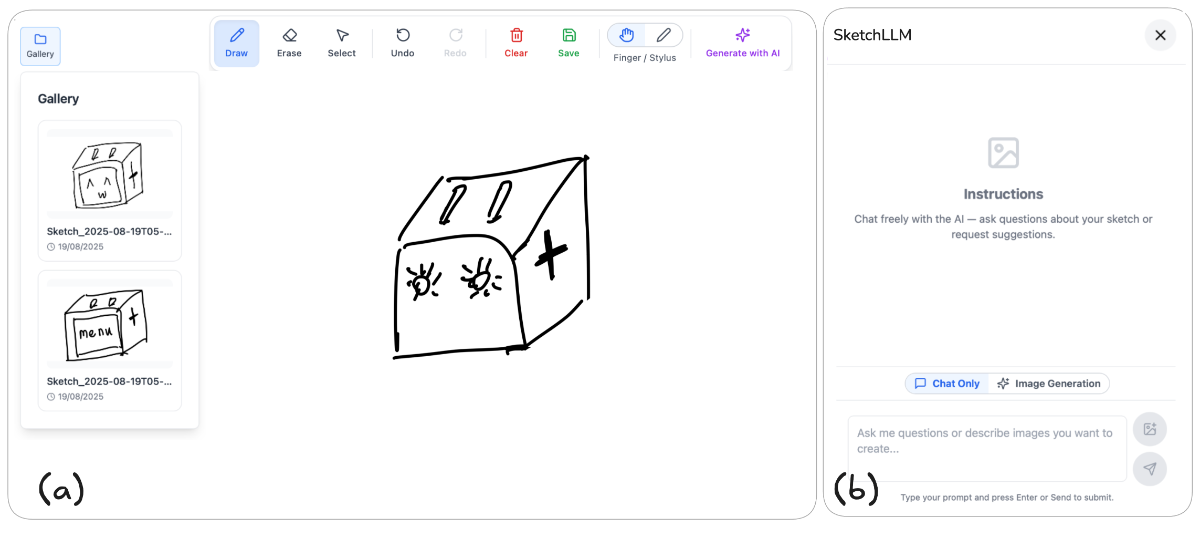}
\caption{
Overview of the \sk{}:
\textit{(a)} \textbf{Sketching module:} Users draw early-stage product concepts (e.g., a toaster) using stylus input on the canvas, which includes a sketch gallery, drawing tools, and controls for launching the \textbf{Multimodal AI Chatbot}.
\textit{(b)} Text- and image-based interaction and generation with the AI for further exploration.
}
\label{fig:ui}
\end{figure*}

As shown in Figure~\ref{fig:ui}, \sk{} consists of two main components that together support sketch-based ideation and multimodal interaction with MLLM.

(a) The \textit{Sketching module} provides a shared canvas for freehand drawing using stylus input, allowing designers to externalise early-stage ideas through sketching.
This canvas supports basic sketching actions such as drawing, erasing, selecting, and revising sketches, and serves as the central workspace.

(b) The \textit{Multimodal AI Chatbot} supports user-initiated interaction with the AI through text and sketch input.
Designers may export selected regions of the sketch canvas into the chatbot as image-based input, and may also import AI-generated images back into the canvas for further sketching and refinement. In addition, designers can submit textual prompts, optionally combined with sketch input, to explore ideas through text-based suggestions or image generation. Specifically, the text-based conversational features are supported by Gemini 2.0 Flash, while image generation is handled by Gemini 2.5 Flash Image\footnote{\url{https://ai.google.dev/gemini-api/docs/models}}. The \textit{Multimodal AI Chatbot} maintains a unified conversation history that persists throughout the entire session and is consistently shared across both models, ensuring coherent context when switching between text and image generation.

Together, these two components allow \sk{} to support multiple forms of interaction with AI, including sketch-based ideation, verbal externalisation during sketching, proactive AI feedback, and flexible exchange between sketches and AI-generated content.
For the purposes of the user study, these components were selectively enabled or disabled to realise the two interaction conditions described in the following section.

\subsection{Study Design}

\subsubsection{Participants}
We recruited 12 participants (P1--P12; 7 female), all of whom had prior exposure to at least one design-related course, which served as our inclusion criterion. Participants ranged in age from 18 to 30 and included non-design students with design training, design students, entry-level designers, and experienced designers. 
Participants were recruited through university channels, local recruitment platforms, and personal networks.
None had prior exposure to the prototype systems or had participated in the earlier formative study.
Each participant received approximately USD~7.8 in compensation for completing the study.
Participants represented a range of experience backgrounds: three non-design students with design training (ND), six design students (DS), one entry-level designer (EL), and two experienced designers (ED).
Participants reported a median sketching confidence of
$Md_{\text{base}} = 4.5$ ($IQR_{\text{base}} = 4.0$--$5.5$) on a 7-point Likert scale and an average of
$\mu = 3.33$ years of design experience ($\sigma = 2.39$).


\textit{\textbf{Procedure.}}
Each session consisted of three phases.

\textit{Phase 1: Introduction (10 min)} 
Each participant received a brief walkthrough of the system’s core functionalities, including sketching on the canvas and interacting with the multimodal chatbot.

\textit{Phase 2: Design Task (30 min)}
In the design task, each participant completed a simple design task: \dquote{freely ideate toaster concepts during a 30-minute session}.
Both interface systems were deployed on an iPad Pro 13" and participants used it with an Apple Pencil Pro to complete all design work. 
The drawing process was recorded from start to finish using screen capture software on a computer connected to the tablet via a data cable. Participants were instructed to ideate freely and explore as many toaster design concepts as possible within the allotted time of 30 minutes.  All interactions with the system were logged.
During the process, other than the given interface, they were also allowed to use external image search consoles, such as Google Images\footnote{https://images.google.com/} and Pinterest\footnote{https://www.pinterest.com/}, to gather inspiration.

\textit{Phase 3: Interview (10 min).}
After completing the design task, participants took part in a 10-minute semi-structured interview that explored their overall design process and their perceptions of how the system supported their ideation.

\textit{\textbf{Data Collection.}}
We collected three types of data during the study. First, participants’ audio was recorded throughout the session using Otter.ai\footnote{https://otter.ai/} and subsequently manually reviewed and corrected for accuracy. Second, full-screen recordings were captured to document participants’ design processes. 
Third, the prototype automatically logged all in-system operations via web-based event logging, capturing detailed interaction events such as sketching actions, and interactions with the multimodal AI components.

\textbf{\textit{Data Analysis.}}
We analysed participants’ interaction behaviours using automatically recorded system logs, with post-task interviews used to support interpretation.
The analysis was conducted in two steps.

First, interaction sequences were extracted automatically from system logs at the ideation-episode level.
Each ideation instance was represented as an ordered sequence of interaction events, including user inputs (e.g., sketching, typed or dictated prompts, sketch export to the chatbot), AI outputs (text and image generation), and auxiliary actions supporting content transfer across system components.
Sequences were generated based on timestamped system events.
All interaction logs were manually checked to verify correct event parsing and mapping to predefined interaction types.

Second, interaction sequences were analysed at the ideation-instance level.
Ideation instances were identified based on participants’ post-task interviews, which were used to segment each participant’s design process into distinct ideas.
For each ideation instance, the corresponding interaction sequence extracted from system logs was identified.

The four interaction modes (\textit{Human-Only}, \textit{Human-Lead}, \textit{AI-Lead}, and \textit{Co-Evolution}) were defined in advance based on prior work on human-centered AI, which characterises human--AI interaction in terms of how responsibility and control are allocated between humans and automation~\cite{shneiderman2022human}.
One researcher then classified each ideation instance into one interaction mode by examining how creative responsibility was distributed between the human and the AI across the associated interaction sequence.

\section{Results}

\begin{figure*}[ht]
    \centering

    \begin{subfigure}[t]{0.48\textwidth}
        \centering
        \includegraphics[width=\textwidth]{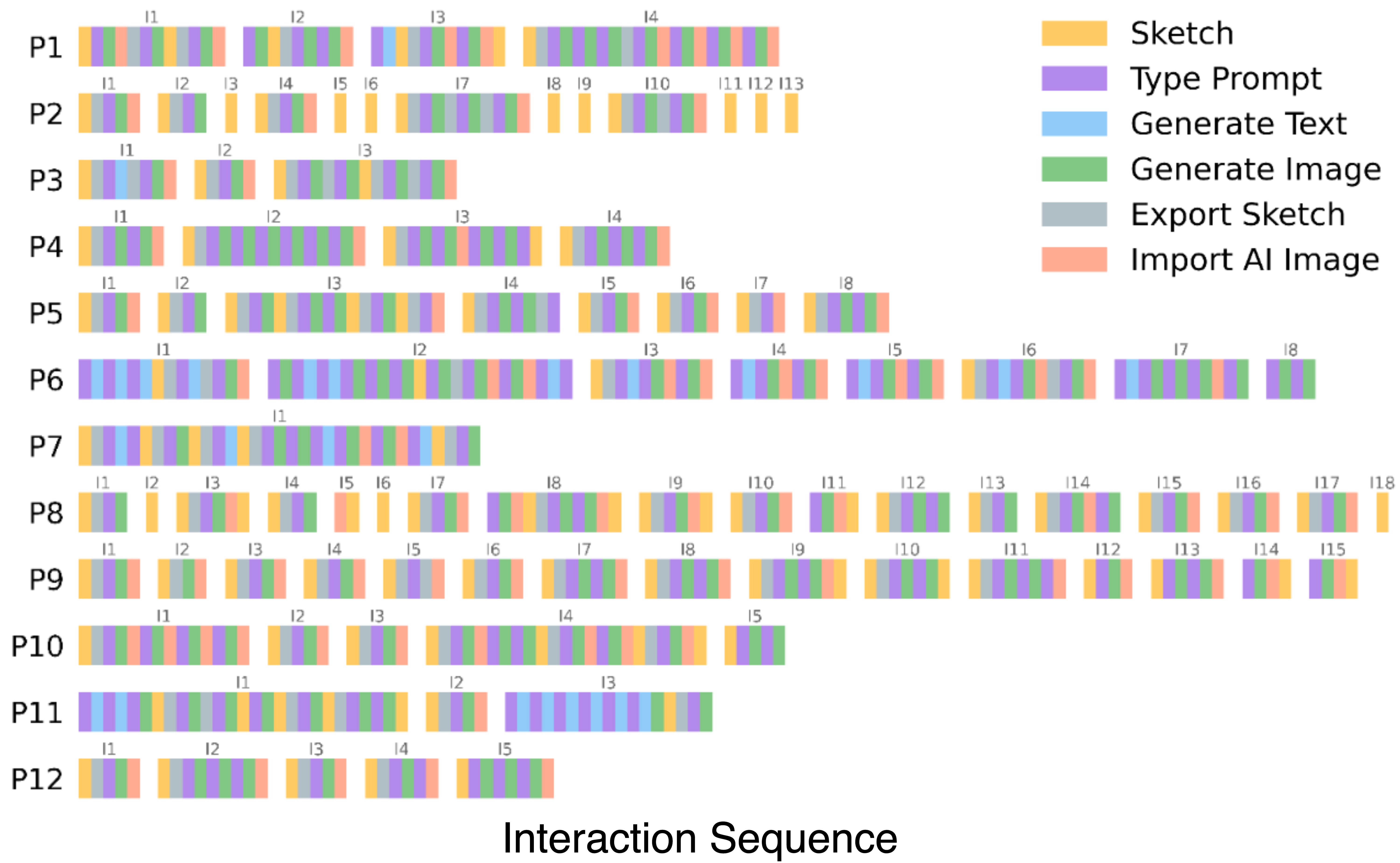}
        \caption{Interaction sequences across participants.}
        \label{fig:individual-a}
    \end{subfigure}
    \hfill
    \begin{subfigure}[t]{0.48\textwidth}
        \centering
        \includegraphics[width=\textwidth]{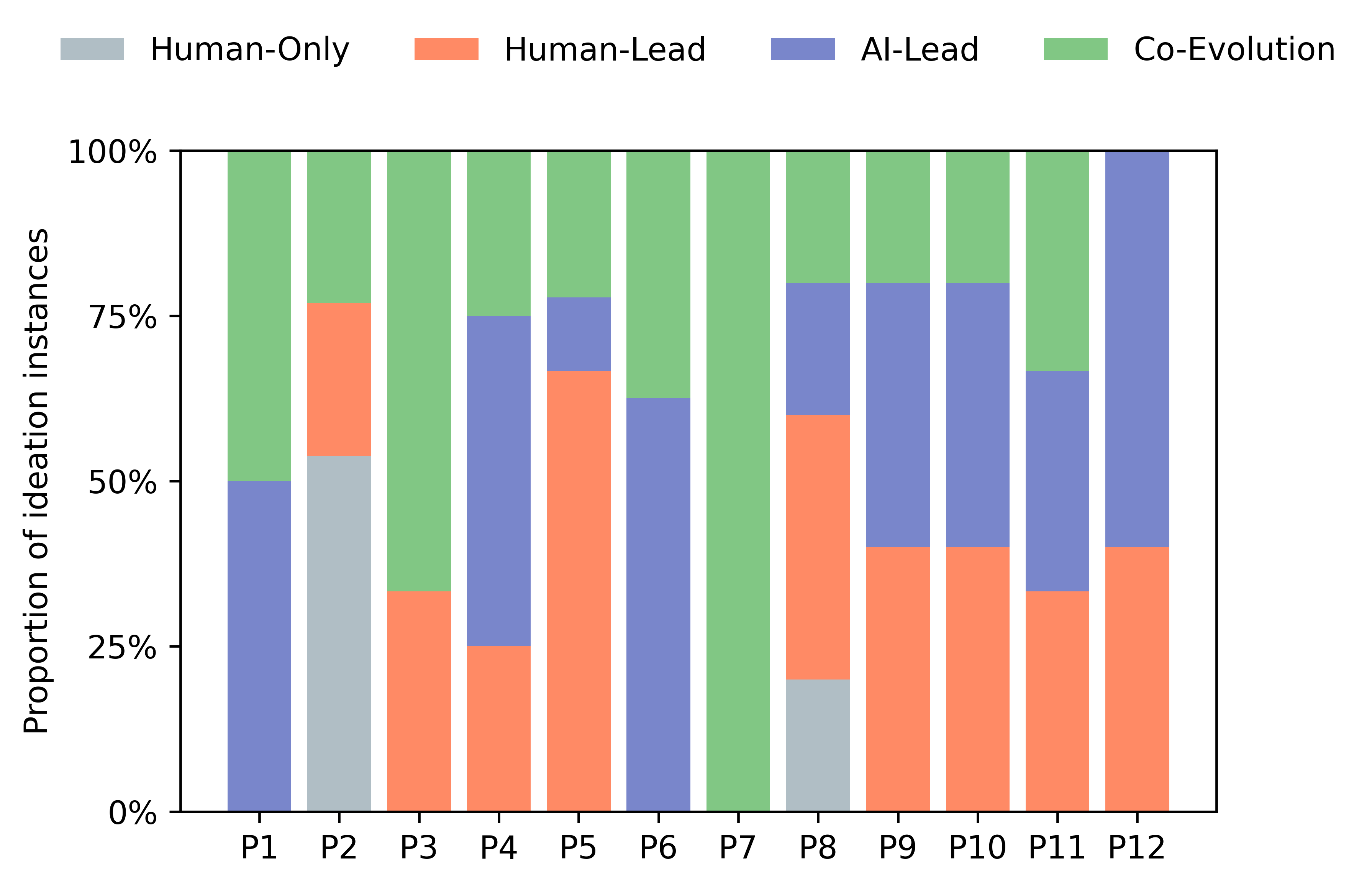}
        \caption{Distribution of interaction regimes per participant.}
        \label{fig:individual-b}
    \end{subfigure}

    \caption{
    \textbf{(a)} Interaction sequences derived from system logs.
    Each horizontal bar represents one participant, with coloured segments indicating the sequence of actions taken for individual ideation instances.
    Action sequences are segmented by ideation episode, with all actions counted equally.
    \textbf{User Input} (\textit{Sketch}, \textit{Type Prompt});
    \textbf{AI Output} (\textit{Generate Text}, \textit{Generate Image});
    and \textbf{Other} (\textit{Export Sketch}, \textit{Import AI Image}).
    \textbf{(b)} Proportional distribution of ideation instances across interaction modes for each participant, normalised to 100\%. Colours indicate \textit{Human-Only}, \textit{Human-Lead}, \textit{AI-Lead}, and \textit{Co-Evolution}.
    }
    \label{fig:individual}
\end{figure*}

Figure~\ref{fig:individual} reports the analysis at two complementary levels.
Figure~\ref{fig:individual}(a) visualises, for each participant, the interaction sequence associated with each ideation instance (i.e., each distinct idea segmented from post-task interviews).
Figure~\ref{fig:individual}(b) aggregates these instances by classifying each ideation instance into one of four interaction modes---\textit{Human-Only}, \textit{Human-Lead}, \textit{AI-Lead}, and \textit{Co-Evolution}---based on how creative responsibility was distributed between the human and the AI across its interaction sequence.
We then computed, for each participant, the proportion of ideation instances falling into each mode.

Averaged across participants (N=12), Co-Evolution accounted for the largest proportion of interactions (34.8\%), followed by AI-Lead (30.6\%) and Human-Lead (28.5\%), while purely Human-Only interactions were rare (6.2\%).

\subsection{Co-Evolution}

The \textit{Co-Evolution} mode represents the most integrated form of human--AI collaboration observed in our study.
Rather than assigning stable roles, participants engaged in iterative cycles in which sketches, prompts, and AI-generated outputs continuously informed one another.
Design responsibility shifted dynamically over time, with neither the human nor the AI consistently occupying a dominant position.

P1 (\textit{Pixel Block Toaster}) demonstrated this dynamic process.
An initial rough sketch was rendered by the AI as a simple grid-like toaster, which in turn prompted P1 to issue follow-up instructions such as \dquote{Make it marketable to kids and add a way to heat the blocks.}
Subsequent AI outputs introduced toy-like aesthetics, leading to further prompts requesting alternative colour schemes.
The final design emerged through successive rounds of sketching, prompting, and AI generation rather than from a single decisive step.
A similar trajectory was observed in P7 (\textit{Smart Toaster Oven}), where an initial sketch was followed by an explicit request for AI critique.
The AI’s suggestions (e.g., \dquote{Sleeker, modern aesthetics}) were then incorporated through additional prompts specifying materials and dimensions.

\textit{Co-Evolution} reflects a tightly coupled interaction in which design ideas emerged through ongoing negotiation between human intent and AI responses.
Ideation unfolded as a process of mutual shaping rather than unilateral control.

\subsection{AI-Lead}

The \textit{AI-Lead} mode represents a further shift in the allocation of creative responsibility, in which participants deliberately delegated form-giving decisions to the AI.
Rather than providing detailed sketches, participants relied on high-level textual descriptions, allowing the AI to determine visual structure, aesthetic style, and object complexity.
Here, the human role shifted from construction toward evaluation and selection.

P6 (\textit{Rocket Launch Station Toaster}) illustrates this pattern.
Starting from a purely textual description of a toaster that could \dquote{toast and fire the cooked toast like a launch station,} the AI generated a rocket-shaped appliance that P6 largely accepted without modification.
As P6 explained, \dquote{I didn’t describe the look... I just want to make sure it make sense.}
A similar dynamic occurred in P6’s \textit{Dragon Head Toaster} concept, where prompts such as \dquote{a monster statue or a dragon head that can flame the toast} led to a detailed sculptural form.
After minor refinements to constrain the concept (e.g., \dquote{No machine, just a dragon head}), P6 allowed the AI to define the object’s visual complexity and style.

\textit{AI-Lead} mode reflects a conscious trade-off: participants relinquished fine-grained control over form in exchange for rapid ideation and visually rich outputs generated by the AI.

\subsection{Human-Lead}

The \textit{Human-Lead} mode captures cases in which participants retained primary control over form and structure, while selectively using AI as a rendering aid.
In this mode, sketches served as the authoritative representation of the design, and AI outputs were valued mainly for enhancing surface-level qualities such as materiality, texture, or visual polish, rather than for generating new structural ideas.

P5 (\textit{Bear \& Fish Toasters}) exemplified this pattern by drawing clear animal-shaped outlines and issuing minimal prompts such as \dquote{Render toaster} or \dquote{Generate.}
The resulting images closely adhered to the sketched contours, adding realistic materials without altering the underlying geometry.
Reflecting on this process, P5 described the AI’s role as \dquote{to bring the picture I drew directly to reality.}
A similar approach was observed in P9 (\textit{F1 Car Toaster}), where a roughly sketched car chassis was paired with a brief prompt requesting a Formula 1–style toaster.
The AI-generated output preserved the original proportions and perspective, leading P9 to remark that they were \dquote{really happy with this one.}

Although participants continued to engage with AI, creative authority remained firmly with the human designer.
Here, AI functioned as a high-fidelity visualiser that enhanced representation, rather than as a source of conceptual direction.

\subsection{Human-Only}

The \textit{Human-Only} mode represents situations in which participants disengaged from AI support and relied exclusively on manual sketching to develop and communicate their ideas.
Rather than reflecting a lack of familiarity with AI tools, this mode typically emerged when participants perceived a mismatch between the AI’s generative outputs and the core mechanical or spatial logic of their design.
In such cases, participants did not attempt to iteratively correct the AI, but instead chose to abandon AI generation altogether.

This pattern is illustrated by P8 (\textit{Conveyor Belt Toaster}), who initially attempted to use AI generation to visualise a \dquote{transparent conveyor belt toaster} in which bread moves horizontally through an enclosed heating system.
Repeated AI outputs failed to capture essential mechanisms, such as a fully enclosed conveyor path or correctly positioned heating elements.
As P8 noted, \dquote{The conveyor is missing.}
After these failures, P8 shifted entirely to producing detailed front and side sketches, using drawing as the primary means to reason about and communicate the system’s internal structure.
Similarly, P2 (\textit{Clock / Weight Toaster}) produced multiple concept sketches without importing them into the AI system, explaining a preference for manual control once the idea was already well-defined: \dquote{I would rather take more time to sketch it out cuz I know exactly what I want already.}

Across these cases, the \textit{Human-Only} mode functioned as a boundary condition for AI use: when participants held a clear internal model or encountered repeated representational breakdowns, AI support was perceived as unnecessary or even obstructive.

\section{Discussion}
Although \textit{Co-Evolution} emerged as the most prevalent interaction mode, our results show that \textit{Human-Lead} and \textit{AI-Lead} interactions together accounted for a substantial proportion of ideation instances.
This indicates that early-stage sketch-based design ideation is not dominated by a single, stable form of human–AI collaboration.
Instead, designers fluidly shift between different modes depending on their moment-to-moment goals, confidence in their ideas, and expectations of the AI’s contribution.

From a human-centred AI perspective, these findings suggest that co-creative design systems should not be optimised solely for tightly coupled co-evolutionary workflows.
Rather, they should support flexible transitions between human-led, AI-led, and co-evolutionary interaction modes, allowing designers to dynamically reallocate creative responsibility as ideation unfolds.
Designing for such adaptability—rather than enforcing a fixed collaboration paradigm—may better align AI systems with the inherently exploratory, heterogeneous, and evolving nature of early-stage design ideation.

\section{Limitations}

Several limitations should be considered when interpreting our findings. 

First, the study was conducted with a relatively small sample and a short-duration, single-type design task. While appropriate for an exploratory poster study, this setup limits the generalisability of the observed interaction patterns. The taxonomy should therefore be understood as a descriptive framework grounded in this specific ideation context rather than a universal account of human–AI collaboration in design.

Second, the classification of interaction modes was conducted by a single researcher. Although guided by a predefined rubric, the absence of a second coder or inter-rater reliability check may limit the robustness of the categorisation, particularly in borderline cases where creative responsibility was less clearly defined.

Third, the four discrete modes simplify what is likely a more continuous and multidimensional allocation of creative responsibility. In practice, initiation, decision-making, and evaluation may be distributed differently across moments of collaboration. Future work could extend this framework using continuous measures or additional dimensions to better capture such nuance.

Finally, our findings are grounded in the currently dominant chatbot-style MLLM paradigm (e.g., ChatGPT- or Gemini-like systems) combined with sketch input. The observed human–MLLM interaction patterns therefore reflect collaboration under turn-based, prompt-driven interfaces. Future work should investigate MLLM systems designed for concurrent sketching and speaking~\cite{shi2025talksketch}, agentic co-creation, and other emerging multimodal interaction paradigms to examine how alternative interface designs shape interaction dynamics.

\begin{acks}
This research is supported by the Ministry of Education, Singapore, under its SUTD Kickstarter Initiative (SKI 2021\_05\_16).
Any opinions, findings and conclusions or recommendations expressed in this material are those of the author(s) and do not reflect the views of the Ministry of Education, Singapore.
\end{acks}

\bibliographystyle{ACM-Reference-Format}
\bibliography{main}

\end{document}